\documentstyle[prl,aps,epsfig,floats,psfig,twocolumn]{revtex}
\tightenlines
\begin{document}
\twocolumn[\hsize\textwidth\columnwidth\hsize\csname
@twocolumnfalse\endcsname
\vglue -0.7cm
\hfill{\vbox{\hbox{CERN-TH/2001-332}
         \hbox{IFIC/01-65}
         \hbox{hep-ph/0111432}
         }}
\title{On the Effect of $\theta_{13}$ on the Determination of
Solar Oscillation Parameters at KamLAND}
\author{M.~C.~Gonzalez-Garcia$^{1,2,3}$ \thanks{concepcion.gonzalez@cern.ch} 
and  C.~Pe\~na-Garay$^{2}$ \thanks{penya@ific.uv.es} }
\vskip 1cm
\address{
  $^1$ Theory Division, CERN, CH-1211 Geneva 23, Switzerland\\
  $^2$Instituto de F\'{\i}sica Corpuscular,  
  Universitat de  Val\`encia -- C.S.I.C.\\
  Edificio Institutos de Paterna, Apt 22085, 46071 Val\`encia, Spain\\
  $^{3}$ C.N. Yang Institute for Theoretical Physics\\
  State University of New York at Stony Brook\\
  Stony Brook,NY 11794-3840, USA\\}
\maketitle
\vskip 1cm
\begin{abstract}
If the solution to the solar neutrino puzzle falls in the LMA region,
KamLAND should be able to measure with good precision 
the corresponding oscillation parameters after a few years of data taking. 
Assuming a positive signal, we study their expected 
sensitivity to the solar parameters ($\theta_{12}$,$\Delta m^2_{21}$)
when considered in the framework of three--neutrino mixing after 
taking into account our ignorance on the mixing angle 
$\theta_{13}$. We find a simple ``scaling'' dependence of the 
reconstructed $\theta_{12}$ range with the value of $\theta_{13}$ 
while the $\Delta m^2_{12}$ range is practically unaffected. 
Our results show that the net effect is approximately equivalent to 
an uncertainty on the overall neutrino flux normalization of up to
$\sim$ 10\%.
\end{abstract}
\vskip2pc]

The Sudbury Neutrino Observatory (SNO) measurement on the charged current 
reaction for solar neutrino absorption in deuterium \cite{sno} has 
provided an important piece of information in the path to solve the solar 
neutrino problem (SNP) \cite{chlorine,sage,gallex,gno,sksol00}. 
In particular, all the post-SNO global analysis 
\cite{our1,lisi,goswami,smirnov,our2} have shown that the inclusion of 
the SNO results have further strengthen the case for solar neutrino 
oscillations with large mixing angles with best fit in the region 
with larger $\Delta m^2$, LMA.  Unfortunately, the LMA region is broad in 
mixing and mass splitting due to the uncertainties in the solar 
fluxes and the lack of detailed data in the low energy range.

This situation will be improved in the very near future, in particular, if LMA 
is the right solution to the SNP. If Nature has arranged things favourably, 
the SNO measurements of the day-night asymmetry and the neutral to charged 
current ratio could help to identify and constrain the 
LMA solution \cite{our2}. Furthermore, the terrestrial experiment, 
KamLAND \cite{KamLAND}, should be able to identify the oscillation 
signal and significantly constrain the region of parameters 
for the LMA solution.

The KamLAND reactor neutrino experiment, which is to start 
taking data very soon, is sensitive to the LMA region of the solar neutrino 
parameter space. After a few years of data taking, it should be capable of 
either excluding the entire LMA region or, not only establishing 
$\nu_{e}\leftrightarrow \nu_{\rm other}$ oscillations, but also 
measuring the oscillation parameters $(\tan^2\theta_{12},\Delta m^2_{21})$ 
with unprecedented precision \cite{Barger_KAM,BS_KAM,SV,MP_KAM,Kam_KAM,dgp}.  
Previous analysis of the attainable accuracy in the determination of 
the oscillation parameters at KamLAND have studied the  
effect of the time dependent fuel composition, the 
knowledge of the flux uncertainty, the role of the geological neutrinos as
well as the combined analysis of solar and KamLAND data.
These studies have been performed in the simplest 
two-neutrino oscillation scheme or equivalently for three-neutrino 
oscillations assuming a small fixed  value of $\theta_{13}$ 
\cite{Barger_KAM,BS_KAM,SV,MP_KAM,Kam_KAM,dgp}.

In this letter, we revisit the problem of how the precision to 
which KamLAND should be able to measure the solar oscillation parameters, 
$\Delta m_{12}^2$ and $\theta_{12}$, is affected by our ignorance
on the exact value of the mixing angle $\theta_{13}$. At present our
most precise information on this parameter comes from the negative 
results from the CHOOZ reactor experiment \cite{CHOOZ}, which, when 
combined with the results from the atmospheric neutrino experiments 
\cite{atmos} results into a 3$\sigma$ upper bound  
$\sin^2\theta_{13}\lesssim 0.06$\cite{total1,total2}. 
To address this question we study how the reconstructed range of 
$\Delta m_{12}^2$ and $\theta_{12}$ depends on the value of the mixing 
angle  $\theta_{13}$.  
We conclude that the reconstructed $\Delta m_{12}^2$ is very mildly 
affected 
by the value of $\theta_{13}$, while the $\theta_{12}$ range scales 
with $\theta_{13}$ in a simple way. We determine the reconstructed region 
of solar parameters obtained from a given signal once $\theta_{13}$
is left free to vary below the present 
bound. We find that the net effect is approximately equivalent to 
that of an uncertainty on the overall neutrino flux normalization of 
up to $\sim$ 10\% and it should be taken into account once enough
statistics is accumulated.


KamLAND is a reactor neutrino experiment located at the old Kamiokande 
site in the Kamioka mine in Japan. 
It is sensitive to the $\bar{\nu}_e$ flux from some 10+ reactors 
which are located ``nearby.'' The distances from the different reactors 
to the experimental site vary from slightly more than 80~km to over 
800~km, while the majority (roughly 80\%) of the neutrinos travel 
from 140~km to 215~km. KamLAND ``sees'' the antineutrinos by
detecting the total energy deposited by recoil positrons, which are
produced via $\bar{\nu}_e+p\rightarrow e^{+}+n$. The total visible 
energy corresponds to 
$E_{e^+}+m_{e}$, where $E_{e^+}$ is the total energy of the positron and
$m_e$ the electron mass. The positron energy, on the other hand, 
is related to the incoming antineutrino energy 
$E_{e^+}=E_{\nu}-1.293$~MeV up to corrections related
to the recoil momentum of the daughter neutron (1.293~MeV is the 
neutron--proton mass difference). KamLAND is expected to measure 
the visible energy  with a resolution which is expected to be better than
$\sigma(E)/E=10\%/\sqrt{E}$, for $E$ in MeV \cite{KamLAND,Kam_KAM}. 

The antineutrino spectrum which is to be measured at KamLAND 
depends on the power output and fuel composition of each reactor 
(both change slightly as a function of time), and on the cross section 
for $\bar{\nu}_e+p\rightarrow e^{+}+n$. For the results
presented here we will follow the flux and the cross section 
calculations and the statistical procedure described in Ref.~\cite{dgp}. 
We use one ``KamLAND-year'' as the amount of time it takes KamLAND to 
see 800 events with visible energy above 1.22~MeV. This is roughly
what is expected after one year of running (assuming a fiducial volume 
of 1 kton), if all reactors run at (constant) $78\%$ of their maximal 
power output \cite{KamLAND}. We assume a constant chemical composition 
for the fuel of all reactors
(explicitly,  53.8\% of $^{235}$U, 32.8\% of $^{239}$Pu, 
7.8\% of $^{238}$U, and
5.6\% of $^{241}$Pu, see \cite{Barger_KAM,chem_comp}). 

The shape of the energy spectrum of the incoming neutrinos can be
derived from a phenomenological 
parametrisation, obtained in \cite{shape_spectrum},
\begin{equation}
\frac{{\rm d}N_{\bar{\nu}_e}}{{\rm d}E_{\nu}}
\propto e^{a_0+a_1E_{\nu}+a_2E^2_{\nu}},  
\label{eq:spectrum}
\end{equation}
where the coefficients $a_i$ depend on the parent nucleus. 
The values of $a_i$ for the different isotopes we used are 
tabulated in \cite{shape_spectrum,MP_KAM}. These 
expressions are very good approximations of the (measured) 
reactor flux for values of $E_{\nu}\gtrsim 2$~MeV. 

The cross section for $\bar{\nu}_e+p\rightarrow e^{+}+n$ has been 
computed including corrections related to the recoil momentum of 
the daughter neutron in \cite{cross_sec}. 
We used the hydrogen/carbon ratio, r=1.87, from the proposed chemical 
mixture (isoparaffin and pseudocumene) \cite{KamLAND}.
It should be noted that the energy spectrum of antineutrinos 
produced at nuclear reactors has been measured with good accuracy 
at previous reactor neutrino experiments 
(see \cite{KamLAND} for references). For this reason, we will first 
assume that the expected (unoscillated) antineutrino energy spectrum 
is known precisely. 
Some of the effects of uncertainties in the incoming flux on 
the determination of oscillation parameters have been studied 
in \cite{MP_KAM}, and are supposedly small.   

In order to simulate events at KamLAND, we need to compute 
the expected energy spectrum for the incoming reactor antineutrinos 
for different values of the neutrino oscillation parameters
mass-squared differences and  mixing angles. In the framework of
three-neutrino mixing the $\nu_e$ survival probability depends on 
the two relevant mass differences and the three-mixing angles but 
it can be simplified if we take into account that:\\
-- Matter effects are completely negligible at KamLAND-like baselines.\\
-- As observed in \cite{BS_KAM}, for 
$\Delta m^2\gtrsim 3\times 10^{-4}$~eV$^2$, 
the determination of $\Delta m^2$ is rather ambiguous. This is due to the 
fact that if $\Delta m^2$ is too large, the KamLAND energy resolution is not 
sufficiently high to resolve the oscillation lengths associated 
with these values of $\Delta m^2$ and $E_{\nu}$. This allows us 
to consider the higher $\Delta m^2_{23}$  and $\Delta m^2_{13}$ 
to be averaged. 

Thus, the relevant (energy dependent) electron antineutrino survival 
probability at KamLAND is
\begin{eqnarray}
P(\bar{\nu}_e\leftrightarrow\bar{\nu}_e) &= &  
\sin^4\theta_{13} + \cos^4\theta_{13} \\\nonumber &&
\left[ 1 - \sum_i f_i \sin^22\theta_{12}\sin^2\left(\frac{1.27\Delta m^2_{21}L_i}{E_{\nu}}\right)\right],
\label{prob}
\end{eqnarray} 
where $L_i$ is the distance of reactor $i$ to KamLAND in $km$, 
$E_{\nu}$ is in GeV and $\Delta m^2_{12}$
is in eV$^2$, while $f_i$ is the fraction of the total neutrino flux which 
comes from reactor $i$ (see \cite{KamLAND}).

From Eq.~(\ref{prob}) we can easily derive the effect of the nonzero 
$\theta_{13}$. The energy independent 
term contains the factor $\sin^4\theta_{13} + \cos^4\theta_{13}$ while the 
energy dependent term contains $\cos^4\theta_{13}$. Thus the shape
of the spectrum (this is, the ratio of the energy
dependent versus the energy independent term) is only modified by
a factor $\frac{\cos^4\theta_{13}}{\sin^4\theta_{13} + \cos^4\theta_{13}}
\sim 1-\sin^4\theta_{13}$. Given the present bound we 
conclude that $\theta_{13}$ does not affect significantly the 
shape of the spectrum which is the most relevant 
information in the determination of $\Delta m^2_{12}$. 
Conversely, the overall spectrum normalization is scaled by 
$\cos^4\theta_{13}\sim 1-2\sin^2\theta_{13}$ and this factor introduces 
an non-negligible effect.

In order to quantify the effect of this term we have simulated 
the KamLAND signal corresponding to some points in the parameter space 
(see Table \ref{range}). Following the approach in Ref.~\cite{dgp}
our simulated data sets are analysed via a 
standard $\chi^2$ function,
\begin{eqnarray}
&&\chi^2(\theta_{12},\Delta m_{12}^2,\theta_{13})
= \nonumber\\
&&\sum_{j=1}^{N_{\rm bin}} 
\frac{\left(N_j(\overline{\theta_{12}},
\overline {\Delta m}_{12}^2,\overline{\theta_{13}}) )
- T_j(\theta_{12},\Delta m_{12}^2,\theta_{13})
\right)^2}{\left(\sqrt{N_j}\right)^2} + N_{\rm d.o.f},\nonumber
\label{eq:chi2}
\end{eqnarray}
where $N_j(\overline{\theta_{12}},
\overline{\Delta m}_{12}^2,\overline{\theta_{13}})$ is the 
number of simulated events in the $j$-th energy bin which would 
correspond to the parameters 
$\overline{\theta_{12}}$ $\overline{\Delta m}_{12}^2$ $
\overline{\theta_{13}}$ 
(see first column in Table \ref{range} 
for the values of the 5 simulated points). 
$T_j(\theta_{12},\Delta m_{12}^2,\theta_{13})$ is the theoretical 
prediction for the number of events in the $j$-th energy bin  
as a function of the oscillation parameters. 
$N_{\rm bin}=12$ is the total
number of bins (binwidth is 0.5 MeV), and the added constant, 
$N_{\rm d.o.f}$, is the number  of  degrees of freedom. This is included 
in order to estimate the statistical capabilities 
of an {\sl average} experiment. An alternative option would  be not to 
include the  $N_{\rm d.o.f}$ term but to include random 
statistical fluctuations in the simulated data as done in 
Ref.{\cite{Barger_KAM}. We have verified that our results
are not quantitatively affected by the choice of simulation procedure.
The fit is first done for visible energies $1.22<E_{vis}<7.22$ MeV.
Note that we assume statistical errors
only, and do not include background induced events. This seems to be a 
reasonable assumption, given that KamLAND is capable of 
tagging the $\bar{\nu}_e$ by looking for a delayed $\gamma$ signal 
due to the absorption of the recoil neutron.  There still 
remains, however,  the possibility of irreducible backgrounds from 
geological neutrinos in the lower energy bins ($E_{vis}\lesssim 2.6$ MeV)
\cite{Kam_KAM,dgp}. To verify how this possible background may 
affect the effect here studied we have repeated the analysis discarding the 
three lower energy bins i.e. considering only events with  
visible energies $2.72<E_{vis}<7.22$ MeV. 

We have generated the signal for the five points in parameter
space listed in Table~\ref{range}. For the sake of concreteness
we have chosen the five points with different values of 
$\overline{\Delta m}_{12}^2$
and $\tan^2\overline{\theta_{12}}$ distributed within the 3$\sigma$
allowed LMA region from the present analysis of the solar data \cite{our2}
and with $\overline{\theta_{13}}=0$. 
For each of the simulated points we obtained the 
reconstructed region of parameters in the plane 
$\Delta m^2_{12}$ $\tan^2\theta_{12}$ by finding the minimum $\chi^2$ and
then calculating the confidence level (CL) for two degrees of freedom 
assuming three KamLAND-years of simulated data. The number of expected
events corresponding to each of the simulated points 
with $1.22<E_{vis}<7.22$ MeV 
is given in Table~\ref{range} 
and in Table~\ref{range2} for $2.72<E_{vis}<7.22$ MeV. 
We have repeated this procedure for the same five simulated signals 
under different assumptions for the value of the reconstructed $\theta_{13}$.

In Fig.~\ref{fig:chi2}(a) we show the allowed regions in the 
$(\tan^2\theta_{12},\Delta m^2_{12})$ plane assuming that we know 
a priory that $\theta_{13}=0$ (which is the simulated value). In other words 
in our minimization procedure we fix $\theta_{13}=0$ in $T_j$ and 
the only fitted parameters
are $\tan^2\theta_{12}$ and $\Delta m^2_{12}$. This case corresponds to 
the usual  two-neutrino analysis. Given our $\chi^2$ prescription the
best fit reconstructed point corresponds exactly with the simulated point. 
The shown regions  correspond to 1$\sigma$, 2$\sigma$ and 3$\sigma$ for 
2 d.o.f ($\Delta\chi^2= 2.30$, 6.18 and 11.83 respectively). 
Similar regions are obtained if 
$\theta_{13}$ if different from zero
but assumed to be known so that its value in the simulated number of events 
and in the reconstructed one is kept to be the same and constant. 
In the fourth column in Table~\ref{range} we list the 3$\sigma$ 
allowed range for $\tan^2\theta_{12}$ for each of the five simulated points.
Table~\ref{range2} contains the corresponding results from the 
``conservative'' analysis in which the  three first bins have been removed.
The main effect is a decrease in the statistics which translates into 
slightly larger ranges. 
We only list the reconstructed range in the first octant but, as shown 
in the figure, due to the negligible 
matter effects, the results  from KamLAND will give us a 
degeneracy in $\theta_{12}$ and an equivalent
range is obtained in the second octact corresponding to 
$\tan^2\theta_{12}\rightarrow 1/\tan^2\theta_{12}$. Solar data will be
able to select the allowed region in the first octant and we show the 
3$\sigma$ contours of the LMA region from the latest analysis.
As discussed in Ref.~\cite{dgp}, if the mixing angle is far enough from 
maximal mixing, the allowed region is clearly separated from the mirror one.

In order to illustrate the effect of the unknown $\theta_{13}$ we have
repeated this exercise but now using a different value of  
$\theta_{13}$ for the simulated point and the reconstructed ones. 
In Fig.~\ref{fig:chi2}(b) we show the reconstructed regions in
$(\tan^2\theta_{12},\Delta m^2_{12})$ corresponding to the same five 
generated points (which are marked by stars in the figure) but using 
$\theta_{13}=12.6^\circ (\sin^2\theta_{13}=0.048)$ in the calculation 
of the expected number of events $T_j$. From the figure we see that 
best fit reconstructed points (which are marked by squares in the figure) 
as well as the allowed regions are shifted in mixing angle 
with respect to the ones in  Fig.~\ref{fig:chi2}(a)
while $\Delta m^2_{12}$ remains practically unaffected.
In tables~\ref{range} and \ref{range2} we list the reconstructed ranges of 
$\tan^2\theta_{23}$ for this academic case.
\begin{figure}
\begin{center}
\psfig{file=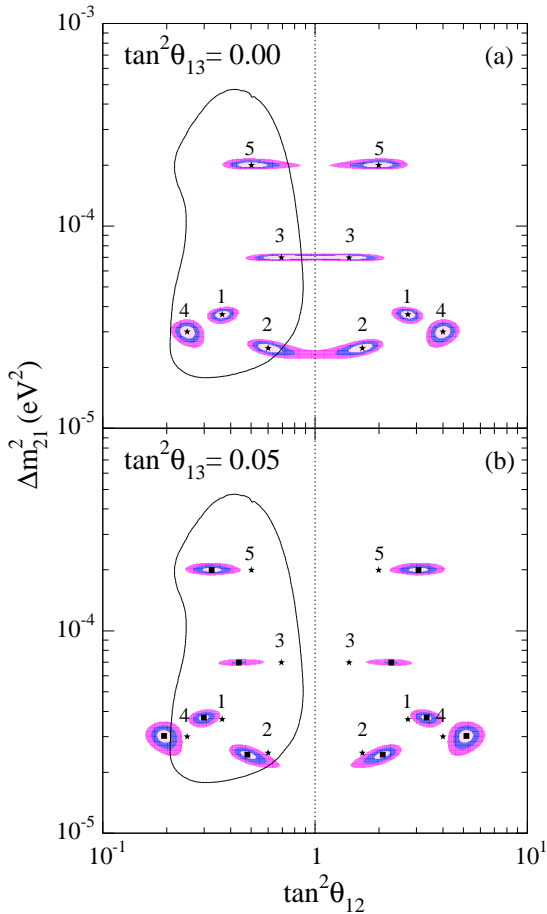,width=0.9\columnwidth}
\end{center}
\caption{(a) Regions of the 
$(\Delta m^2_{21}\times\tan^2\theta_{12})$-parameter space 
allowed by three KamLAND-years of simulated data at the  
1$\sigma$, 2$\sigma$ and 3$\sigma$ CL, 
for different input values of $\overline{\Delta m}^2_{21}$ and 
$\tan^2\overline{\theta_{12}}$ ($\overline{\theta_{13}}=0$) 
and assuming $\theta_{13}=0$. 
The stars indicate the best fit points (corresponding also to the 
simulated signals).
(b) Regions of the $(\Delta m^2_{21}\times\tan^2\theta_{12})$-parameter 
space allowed by three KamLAND-years of simulated data 
at the 1$\sigma$, 2$\sigma$ and 3$\sigma$ CL
for $\theta_{13}=12.6^\circ (\sin^2\theta_{13}=0.048)$.
The stars indicate the points used to simulate the signal 
(with fixed $\overline{\theta_{13}}=0$) while the squares
 indicate the reconstructed best fit point 
(with fixed $\theta_{13}=12.6^\circ$).}
\label{fig:chi2}
\end{figure}

The observed shift can be easily understood as follows. From Eq.~(\ref{prob}), 
the total number of events is equal for different $\theta_{13}$ with the 
condition 
\begin{equation}
1-\alpha\times\sin^22\theta_{12}=\cos^4\theta_{13} (1-\alpha\times\sin^22\theta'_{12})
\,,
\end{equation} 
where $\alpha$ is a number coming from all the detailed integration of the
oscillating phase factor which depends mainly on $\Delta m^2_{12}$ and 
which, for completeness, we also list in Table~\ref{range}. For example for 
the simulated point (1), $\tan^2\overline{\theta_{12}}=0.37$, 
$\overline{\Delta m}_{12}^2=3.7\times 10^{-5}$ eV$^2$ and 
$\overline{\theta_{13}}=0$,  
the 3$\sigma$ reconstructed range, $0.31\leq\tan^2\theta_{12}\leq 0.43$, 
for $\theta_{13}=0$ is shifted using the above 
relation with $\alpha=0.65$ to $0.25\leq\tan^2\theta_{12}\leq 0.35$ 
for  $\tan^2\theta_{13}=0.05$ which precisely coincides with the 
values listed in the 5th column of table~\ref{range}. 
Strictly speaking this scaling is slightly
violated due to the change of the spectral
shape with the change from $\theta_{12}$ to $\theta'_{12}$ which
also worsens the $\chi^2_{min}$ for the reconstructed point.

Let us finally consider which are the allowed regions in the parameter space 
$(\tan^2\theta_{12},\Delta m^2_{21})$ taking into account that we just 
know that $\theta_{13}$ is below some limit. In order to do so one must  
integrate over $\theta_{13}$ in the allowed range, or what is 
equivalent, for each pair $(\tan^2\theta_{12},\Delta m^2_{21})$
we must minimize 
$\chi^2(\theta_{12},\Delta m^2_{21},\tan^2\theta_{13})$ with respect to
$\theta_{13}$ (restricted to be below the present bound). 
Notice that, below the bound, we used a flat probability  
distribution for $\theta_{13}$ to keep the analysis just KamLAND-dependent. 
In the future, the combined analysis of solar, atmospheric and CHOOZ 
results with KamLAMD data will allow us to include the 
probability distribution for $\theta_{13}$. 
In Fig.~\ref{fig:chi2min} we shown the allowed regions for such  
$\theta_{13}$-free  analysis, where free means allowed to 
vary below its 3$\sigma$ limit, $\theta_{13}\leq 13.8^\circ$ 
($\sin^2\theta_{13}\leq 0.057)$). As seen by comparing 
Fig.~\ref{fig:chi2min} with  Fig.~\ref{fig:chi2} 
the reconstructed regions are enlarged and roughly correspond to the overlap 
of the allowed regions for the different fixed values of $\theta_{13}$. 
Given our signal generating procedure the best fit reconstructed point 
corresponds to the simulated point in 
$(\tan^2\theta_{12}$, $\Delta m^2_{21})$ and $\theta_{13}=0$.   
The reconstructed ranges can be 
read from the last column in Tables~\ref{range} and Tables~\ref{range2}.
Also comparing the results in both tables one sees that the presence of
this effect is not quantitatively affected by not including in the fit 
the data from the lowest energy bins. The main effect being a decrease 
in the statistics which translates into slightly larger ranges. 
\begin{figure}
\begin{center}
\psfig{file=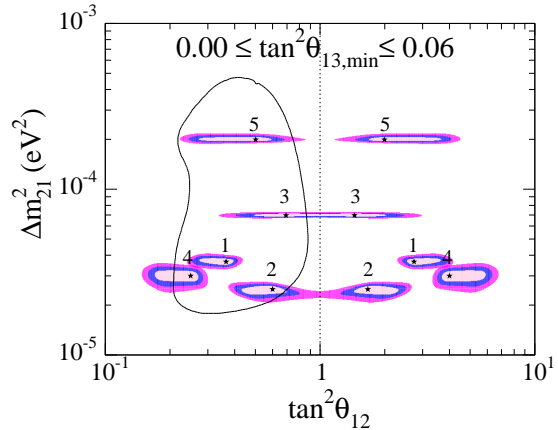,width=0.9\columnwidth}
\end{center}
\caption{1$\sigma$, 2$\sigma$ and 3$\sigma$ CL allowed 
regions of the 
$(\Delta m^2_{21}\times\tan^2\theta_{12})$-parameter space when 
$\theta_{13}$ is left free to vary in the allowed range 
($\theta_{13}<13.8^\circ (\sin^2\theta_{13}=0.057)$).  
The regions are obtained for three KamLAND-years of simulated data and 
for the same five simulated signals as Fig.1(a).
The stars indicate the best fit points
(corresponding also to the simulated points).}
\label{fig:chi2min}
\end{figure}

\begin{table*} 
\begin{center} 
\begin{tabular}{|c|c|c|c|c|c|} 
Signal $\equiv$ ($\tan^2\overline{\theta_{12}},
\overline{\Delta m}^2_{21},\overline{\theta_{13}}$) 
& $N_{ev}$ &$\alpha$ & ~$\theta_{13}=0$ & $\theta_{13}=12.6^\circ (0.048)$ 
& $\theta_{13}<13.8^\circ (0.057)$\\ 
\hline  
~~~~ 1 $\equiv$ (0.37,3.7$\times 10^{-5}$,0)  & 1022
&0.65 & [0.31,0.43]  
&  [0.25,0.35] &  [0.24,0.43] \\
~~~~ 2 $\equiv$ (0.60,2.5$\times 10^{-5}$,0) &802 
&0.65 & [0.48,1.00] 
&  [0.39,0.68] &  [0.38,1.00] \\
~~~~ 3 $\equiv$ (0.70,7.0$\times 10^{-5}$,0) &1335 
&0.56 & [0.47,1.00] 
&  [0.35,0.57] &  [0.34,1.00] \\ 
~~~~ 4 $\equiv$ (0.25,3.0$\times 10^{-5}$,0) & 1237 
&0.65 & [0.21,0.30] 
&  [0.16,0.24] &  [0.15,0.30] \\ 
~~~~ 5 $\equiv$ (0.50,2.0$\times 10^{-4}$,0) & 1321
&0.44 & [0.37,0.85] 
&  [0.25,0.44] &  [0.22,0.85] 
\end{tabular} 
\end{center} 
\vglue .1cm 
\caption{Reconstructed ranges for $\tan^2\theta_{12}$ at 3 $\sigma$
(in the first octant) for 
different $\theta_{13}$ ($\sin^2\theta_{13}$ in parenthesis) cases
for the simulated points listed in the first column. See text for details.}
\label{range} 
\end{table*} 
\begin{table*} 
\begin{center} 
\begin{tabular}{|c|c|c|c|c|c|} 
Signal $\equiv$ ($\tan^2\overline{\theta_{12}},
\overline{\Delta m}^2_{21},\overline{\theta_{13}}$) 
& $N_{ev}$ &$\alpha$ & ~$\theta_{13}=0$ & $\theta_{13}=12.6^\circ (0.048)$ 
& $\theta_{13}<13.8^\circ (0.057)$\\ 
\hline

~~~~ 1 $\equiv$ (0.37,3.7$\times 10^{-5}$,0)  & 756
&0.65 & [0.31,0.44]
&  [0.25,0.36] &  [0.24,0.44] \\
~~~~ 2 $\equiv$ (0.60,2.5$\times 10^{-5}$,0) & 695
&0.72 & [0.41,1.00]
&  [0.35,1.00] &  [0.34,1.00] \\
~~~~ 3 $\equiv$ (0.70,7.0$\times 10^{-5}$,0) & 1167
&0.52 & [0.44,1.00]
&  [0.32,0.59] &  [0.31,1.00] \\
~~~~ 4 $\equiv$ (0.25,3.0$\times 10^{-5}$,0) & 1004
&0.65 & [0.20,0.42]
&  [0.15,0.68] &  [0.14,1.00] \\
~~~~ 5 $\equiv$ (0.50,2.0$\times 10^{-4}$,0) & 1090
&0.44 & [0.36,1.00]
&  [0.24,0.46] &  [0.22,1.00]
\end{tabular} 
\vglue .1cm 
\caption{Same as table 1 after removing the lower energy bins from the fit
(therefore including only events with $2.72<E_{vis}<7.22$ MeV).}
\label{range2}
\end{center}
\end{table*}
We have studied the role of $\theta_{13}$ in the hypothetical case of 
perfect knowledge of the overall flux normalization and fuel composition.
Comparing our results with the expected degradation on the parameter 
determination associated with the uncertainty on those two assumptions 
~\cite{MP_KAM}, we note that the reconstructed ranges obtained 
in the $\theta_{13}$-free  case are very close to those  
obtained in the analysis with perfect knowledge of 
$\theta_{13}$ but where the overall flux normalization is unknown 
by about $\sim$ 10\%. As mentioned above, the role of $\theta_{13}$ 
is essentially the change of the normalization. This is a larger
normalization error that the 3\% expected one in the theoretical calculation
of the flux from the reactors 
(induced from the $\beta$-spectroscopy experiment at the Goesgen reactor 
\cite{gosgen}). We also find that the uncertainty associated with 
$\theta_{13}$ has a larger impact on the 
determination of the mixing angle $\theta_{12}$ than the expected 
error associated with the fuel composition although, unlike this last one, 
it does not affect the determination of $\Delta m^2_{12}$.

Let us point out, that obviously, in order for this effect to become relevant 
it has to be  larger than the expected statistical uncertainty on the
overall flux normalization.
If we repeat this exercise assuming only one KamLAND-years of simulated data,
we find very little difference between the results corresponding
to fixed $\theta_{13}=0$ and the free-$\theta_{13}$ analysis, as expected
since the expected number of events would be of the order $\sim 400$ and
the associated statistical uncertainty for the overall normalization
would be comparable with the maximum effect associated to $\theta_{13}$. 

Summarizing, the KamLAND reactor neutrino experiment, which is to start 
taking data very soon, is sensitive to the LMA region and
should be able of measuring the solar oscillation 
parameters $(\tan^2\theta_{12},\Delta m^2_{21})$ with unprecedented
precision. 
In this letter, we have addressed the question of the degradation
on the determination of the oscillation parameters associated with 
our ignorance of the exact value of $\theta_{13}$.
We have shown that the determination of $\Delta m_{12}^2$ 
is practically unaffected  because the effect of $\theta_{13}$ on 
the shape of the spectrum is very small. 
The dominant effect is a shift in the overall
flux normalization which implies that the reconstructed $\theta_{12}$ 
range scales with $\theta_{13}$ in a simple way. As a consequence 
the allowed region of solar parameters obtained from KamLAND
signal will be broader in $\theta_{12}$. Comparing this effect
with the ones from the expected uncertainties associated with 
the theoretical error on the overall flux normalization and the 
fuel composition, we find that, after enough statistics is accumulated,
the uncertainty associated with 
$\theta_{13}$  may become the dominant source of degradation 
in the determination of the mixing angle  $\theta_{12}$ at KamLAND.

\acknowledgments
We thank A. de Gouvea for comments and suggestions.
MCG-G is supported by the European Union Marie-Curie fellowship
HPMF-CT-2000-00516.
This work was also supported by the Spanish DGICYT under grants PB98-0693
and PB97-1261, by the Generalitat Valenciana under grant
GV99-3-1-01, by the  European Commission RTN network 
HPRN-CT-2000-00148 and by the European Science
Foundation network grant N.~86.

\end{document}